\documentclass{article}

\usepackage{arxiv}

\usepackage[utf8]{inputenc} 
\usepackage[T1]{fontenc}    
\usepackage{hyperref}       
\usepackage{url}            
\usepackage{booktabs}       
\usepackage{amsfonts}       
\usepackage{amsmath}
\usepackage{algorithm}
\usepackage{algpseudocode}
\usepackage{nicefrac}       
\usepackage{microtype}      
\usepackage{lipsum}		
\usepackage{graphicx}
\usepackage{subcaption}
\usepackage{natbib}
\usepackage{doi}
\usepackage{color}

\title{FengWu-4DVar: Coupling the Data-driven Weather Forecasting Model with 4D Variational Assimilation}

\author{
        Yi Xiao $^{1,2}$ \And
	Lei Bai $^{2,\dagger}$ \And
        Wei Xue $^{1,\dagger}$ \And
        Kang Chen $^{2}$ \And
        Tao Han $^{2}$ \And
        Wanli Ouyang $^{2}$ \AND
        $^{\dagger}$ Corresponding author, email to bailei@pjlab.org.cn and xuewei@tsinghua.edu.cn
        \AND
        $^{1}$ Tsinghua University \And 
        $^{2}$ Shanghai Artificial Intelligence Laboratory \And 
}

\renewcommand{\shorttitle}{FengWu-4DVar}

\hypersetup{
pdftitle={A template for the arxiv style},
pdfsubject={q-bio.NC, q-bio.QM},
pdfauthor={David S.~Hippocampus, Elias D.~Striatum},
pdfkeywords={First keyword, Second keyword, More},
}

\begin{document}
\maketitle

\begin{abstract}
Weather forecasting is a crucial yet highly challenging task. With the maturity of Artificial Intelligence (AI), the emergence of data-driven weather forecasting models has opened up a new paradigm for the development of weather forecasting systems. Despite the significant successes that have been achieved (e.g., surpassing advanced traditional physical models for global medium-range forecasting), existing data-driven weather forecasting models still rely on the analysis fields generated by the traditional assimilation and forecasting system, which hampers the significance of data-driven weather forecasting models regarding both computational cost and forecasting accuracy. 
In this work, we explore the possibility of coupling the data-driven weather forecasting model with data assimilation by integrating the global AI weather forecasting model, FengWu, with one of the most popular assimilation algorithms, Four-Dimensional Variational (4DVar) assimilation, and develop an AI-based cyclic weather forecasting system, FengWu-4DVar. 
FengWu-4DVar can incorporate observational data into the data-driven weather forecasting model and consider the temporal evolution of atmospheric dynamics to obtain accurate analysis fields for making predictions in a cycling manner without the help of physical models. 
Owning to the auto-differentiation ability of deep learning models, FengWu-4DVar eliminates the need of developing the cumbersome adjoint model, which is usually required in the traditional implementation of the 4DVar algorithm. 
Experiments on the simulated observational dataset demonstrate that FengWu-4DVar is capable of generating reasonable analysis fields for making accurate and efficient iterative predictions.


\end{abstract}

\keywords{AI forecasting model\and Data assimilation \and auto-differentiation}

\section{Introduction}
\label{sec:introduction}
Weather forecasting is the cornerstone of human society and profoundly impacts all aspects of our daily lives and economic activities. The capacity to forecast weather conditions in advance offers valuable insights that can significantly impact decision-making across various sectors, including agricultural planning, renewable energy generation, and disaster preparedness.

Although the field of weather forecasting is undeniably crucial, it encounters significant challenges. Traditional numerical weather forecasting methods rely on building differential equations based on physical rules and solving these equations for accurate predictions~\citep{kalnay2003atmospheric}. However, the complexity of physical processes like clouds and convection makes medium-term forecasts less accurate~\citep{bauer2015quiet}. Moreover, the computational complexity of solving these differential equations is high, requiring significant investment in supercomputing clusters for weather forecasting. In the past few years, a multitude of Artificial Intelligence (AI) weather forecasting models have emerged as a promising alternative, such as FourCastNet~\citep{pathak2022fourcastnet}, Pangu Weather~\citep{bi2023accurate}, GraphCast~\citep{lam2022graphcast}, FengWu~\citep{chen2023fengwu}, FuXi~\citep{chen2023fuxi}, etc. These models are data-driven, and their forecast accuracy rivals or even surpasses traditional methods like Integrated Forecasting System (IFS) developed by European Centre for Medium-Range Weather Forecasts (ECMWF). Notably, they exhibit forecasting efficiency orders of magnitude higher than that of traditional algorithms. 


Despite the significant achievements of AI forecasting models, there is currently limited attention to the acquisition of initial values for these models. In previous studies, the establishment of initial values frequently depends on the analysis field derived of assimilating data from traditional assimilation and forecasting systems. This reliance on traditional model-derived initial values limits the application scenarios, weakens the efficiency benefits, and constrains the accuracy upper bound of data-driven weather forecasting models in real-world scenarios. This work aims to explore the possibility of eliminating the reliance of data-driven weather forecasting models on physical forecasting models during operational forecasting. 

\begin{figure}[htpb]
\centering
\includegraphics[width=1.0\linewidth]{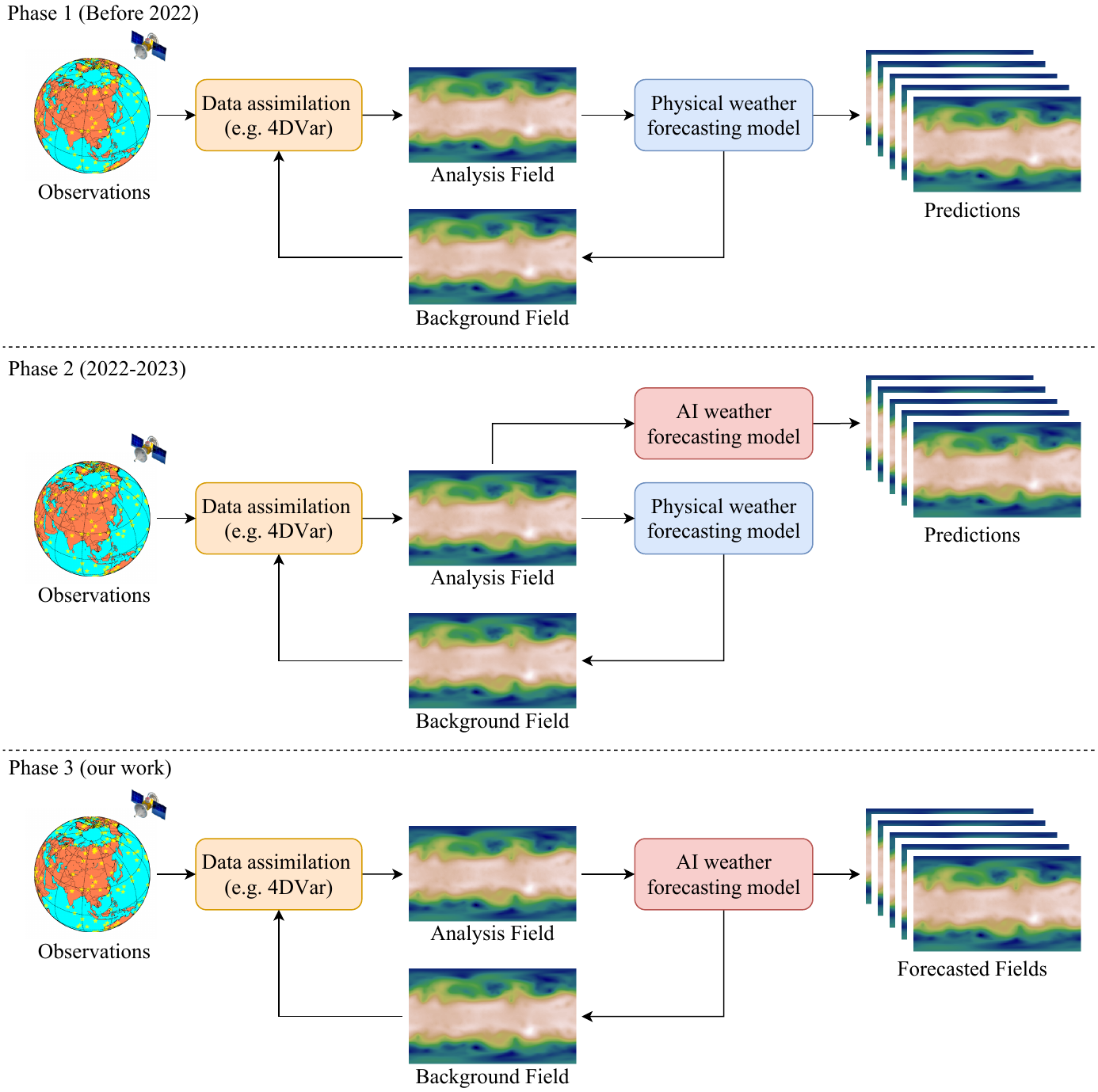}
\caption{\textbf{Three development phases of cyclic weather forecasting systems.}}
\label{fig:pipeline}
\end{figure}

Specifically, we present FengWu-4DVar, a cyclic AI forecasting system coupling FengWu, a data-driven weather forecasting model, with the Four-Dimensional Variational (4DVar) assimilation~\citep{rabier1998extended} algorithm, one of the most sophisticated data assimilation techniques that has been widely adopted in weather forecasting agencies worldwide like ECMWF~\citep{rabier2000ecmwf}. By utilizing observations to continuously correct the predicted results of the AI forecasting model, FengWu-4DVar is capable of performing cyclic forecasting in a self-contained manner. Our forecasting system brings about a twofold contribution, covering both weather forecasting and data assimilation.

As for weather forecasting, we have ushered in a new phase by pioneering independent cyclic forecasting using AI. The development of weather forecasting has undergone two phases, as illustrated in Figure~\ref{fig:pipeline}. 
The first phase features the traditional cyclic forecasting system in which the physical model and data assimilation cooperate to generate the analysis field as the initial value of the forecast, and then the physical model performs weather forecasting by doing integrations from the analysis field.
Since 2022, various data-driven AI models have been developed to substitute the physical weather forecasting model for prediction tasks, marking the second phase of weather forecasting. However, in this phase, the analysis field as the initial value of the integration still needs to be obtained by assimilating the background field generated by the physical model. This results in the forecast results being constrained by the physical model that serves to provide the initial value. 
FengWu-4DVar starts a new phase in which the AI forecasting model is not only leveraged for weather predictions, but also combined with the 4DVar algorithm for providing an analysis field as the initial state of AI weather forecasting. This enables us to eliminate the dependence of data-driven weather forecasting models on physical forecasting models in the course of operational cyclic forecasting.

Our work also contributes to the data assimilation community. The key challenge of implementing the 4DVar algorithm lies in the development of the so-called \textit{adjoint model}~\citep{rabier1998extended}. The adjoint model, roughly speaking, is the conjugate process of the forward model, which determines how the observations at future moments propagate back to the current moment and affect the estimate of the current state. As for the physical weather forecasting models, the adjoint models are coded manually, which requires a substantial engineering effort, and the computational cost of calculating the adjoint models is high. There are at least two advantages of combining AI weather forecasting models and the 4DVar algorithm. The first is that owing to the auto-differentiation ability of deep learning models, FengWu-4DVar is able to calculate adjoint models in an implicit manner, removing the necessity of manually coding the adjoint models, thus saving numerous engineering efforts. The second advantage is related to the computational cost. As has been demonstrated in prior works~\citep{chen2023fengwu, pathak2022fourcastnet}, the computational cost of AI weather forecasting models is much smaller than that of the physical forecasting models. Provided that the computational cost of the adjoint model is proportional to that of the forward model, the 4D-Var algorithm implemented based on the AI weather forecasting model, FengWu, exhibits improved computational efficiency compared to that based on physical models.

We evaluate FengWu-4DVar with simulated observations to verify the effectiveness of our system. Under the condition that the number of observations is only 15\% of the number of grid points, FengWu-4DVar is capable of making stable and accurate cyclic predictions for at least one year. At the same time, FengWu-4DVar is also efficient and can assimilate observations within a 6-hour window in less than half a minute on only one Nvidia A100 GPU card.

\section{Related Work}

In recent years, there have been endeavors to combine data assimilation with AI forecasting models. These works can be generally divided into two categories. 

The first category is dedicated to utilizing AI techniques for accelerating the implementation of data assimilation algorithms. For example, in \citet{fablet2021learning}, an AI forecasting model and an optimization algorithm for 4DVar objective function are jointly learned by two distinct neural networks, and the assimilation efficiency is shown to be significantly improved compared with the traditional method on Lorenz systems. \citet{fablet2021learning} designs an inverse observation operator to learn the mapping from observational data to physical states. Such an operator can better initialize the 4DVar optimization problem and reformulate the objective function in better-behaved physics space instead of observation space, thus accelerating the convergence speed of 4DVar optimization. In \cite{mack2020attention, peyron2021latent, melinc2023neural}, an autoencoder or a VAE is trained to map fields from high-dimensional physical space into the low-dimensional latent space and do assimilation in the latent space in order to reduce the computational cost. 
Despite the success of these endeavors, the coupling of data assimilation algorithms with AI models for making cyclic predictions has not been extensively explored.

A few works aim to couple data assimilation with AI forecasting models to develop a stable and efficient cyclic forecasting system, which aligns with the target of our work. For example, in \citet{hatfield2021building}, a neural network is trained to emulate the non-orographic gravity wave drag parametrization scheme, and the adjoint model is directly derived from neural networks for 4DVar optimization, proving the feasibility of combining AI parameterization scheme with 4DVar. 
In \citet{brajard2020combining, farchi2021using, arcucci2021deep}, analysis fields and AI forecasting models are simultaneously optimized through data assimilation to achieve accurate and stable cyclic forecasting. 
Despite these achievements, the AI model in these works either only serves as a surrogate for part of real-world atmospheric systems or serves to substitute a simplified dynamic system, such as the Lorenz system~\citep{lorenz1996predictability} and the double integral mass dot system. 
However, neural networks for real atmospheric systems are normally much more complex than those for simplified systems.
To the best of our knowledge, integrating AI-based global weather forecasting models for complete real atmospheric systems with the 4DVar algorithm has not been explored, and we are the first to achieve this. 

\section{Four-dimensional Variational Assimilation}
\label{sec:preliminaries}

Denote $\mathbf{x}_t$ the physical states and $\mathbf{y}_t$ the observations at time $t$. Then, the 4DVar algorithm estimates the optimal physical state at time $t=0$ by minimizing the following objective function:
\begin{equation}
\begin{aligned}
    J(\mathbf{x}_0) &= \frac12 \left(\mathbf{x}_0 - \mathbf{x}^b\right)^T \mathbf{B}^{-1} \left(\mathbf{x}_0 - \mathbf{x}^b\right) + \frac12 \sum_{\tau=0}^{T-1} \left(\mathcal{H}\left(\mathbf{x}_\tau\right) - \mathbf{y}_\tau\right)^\text{T} \mathbf{R}_\tau^{-1} \left(\mathcal{H}\left(\mathbf{x}_\tau\right) - \mathbf{y}_\tau\right)  \\
    \mathbf{x}_\tau &= \mathcal{M}_{0\to \tau}(\mathbf{x}_0)
\end{aligned}
\label{equ:4dvar}
\end{equation}

The objective is to compute a maximum likelihood estimate for the initial state $\mathbf{x}_0$ of a trajectory $(\mathbf{x}_0, \ldots, \mathbf{x}_{T-1})$ evolved through a physical model $\mathcal{M}$, given a sequence of observations $\{\mathbf{y}_\tau\}_{\tau=0}^{T-1}$ and a prior estimate $\mathbf{x}^b$. We note here that the subscript $0\to \tau$ of $\mathcal{M}$ stands for integration from time $0$ to time $\tau$. Since we only consider \textit{autonomous} systems~\citep{strogatz2018nonlinear} in this paper, $\mathcal{M}_{0\to \tau} = \mathcal{M}_{t\to t+\tau}$ holds for any $t$, thus we may also rewrite it as $\mathcal{M}_\tau$. The observation operator $\mathcal{H}$ maps physical states into the observation space. For example, physical fields are often modeled on a regular grid, while the positions of observation stations are typically distributed irregularly. Observation operators can map the values of the physical field at regular grid points to the positions of observation stations. The loss function $J(\mathbf{x}_0)$ characterizes the initial condition and the conditional distribution of observations as multivariate normal distributions. The first term incorporates a guess for the initial state $\mathbf{x}_0$ (referred to as the background field $\mathbf{x}^b$), where $\mathbf{B}$ is the background covariance matrix representing the uncertainty associated with this assumption. The second term incorporates the observations, the error variance of which is represented by the matrix $\mathbf{R}_i$.

\paragraph{Function Optimization} 4DVar minimizes objective function via gradient based optimization  like fixed-step gradient descent and L-BFGS~\citep{jorge2006numerical}. The gradient of $J(\mathbf{x}_0)$ can be formulated as
\begin{equation}
    \begin{aligned}
        \frac{\partial J}{\partial \mathbf{x}_0} = \mathbf{B}^{-1} \left(\mathbf{x}_0 - \mathbf{x}^b\right) + \sum_{\tau=0}^{T-1} \mathbf{M}_{\tau\to 0}^\text{T}\mathbf{H}^\text{T}\mathbf{R}_\tau^{-1} \left(\mathcal{H}\left(\mathbf{x}_\tau\right) - \mathbf{y}_\tau\right),
    \end{aligned}
\end{equation}
where $\mathbf{M}_{\tau\to 0}^\text{T} = \left(\frac{\partial \mathcal{M}_{0\to \tau}(\mathbf{x})}{\partial \mathbf{x}}\right)^\text{T}$ is the adjoint model~\citep{rabier2003variational} of $\mathcal{M}_{0\to \tau}$ and $\mathbf{H} = \frac{\partial \mathcal{H}(\mathbf{x})}{\partial \mathbf{x}}$ is the linearized observation operator. 

For conventional numerical models, constructing adjoint models $\mathbf{M}_{\tau\to 0}^\text{T}$ requires a tremendous amount of engineering effort, and each development of a new model necessitates the re-coding of the adjoint model~\citep{tr2006accounting}. Furthermore, even after the completion of adjoint model construction, due to the high computational complexity of the forward model, the computational burden of the adjoint model is substantial. This significant computational load is further multiplied in optimization algorithms, as iterative optimization algorithms demand multiple gradient calculations. 

\paragraph{Cyclic Forecasting} In operational weather forecasting centers, establishing a self-contained forecasting framework is achievable through the integration of data assimilation and model forecasts. Specifically, denoting $J\left(\cdot\middle|\mathbf{x}^b, \{\mathbf{y}_\tau\}_{\tau=0}^{T-1}\right)$ the objective function with respect to observations $\{\mathbf{y}_\tau\}_{\tau=0}^{T-1}$ and $\mathbf{x}^b$ background field, the synergy of the 4DVar algorithm with model forecasts is implemented as shown in Algorithm~\ref{alg:cyclic}, where the output $\{\mathbf{x}_{lT}^b\}_{l=0}^L$ and $\{\mathbf{x}_{lT}^a\}_{l=0}^L$ represent the sequences of background fields and analysis fields, respectively.
\begin{algorithm}[ht]
\caption{Cyclic Forecasting with 4DVar}\label{alg:cyclic}
\begin{algorithmic}
\Require: Prior estimate of the initial state $\mathbf{x}_0^b$, physical model $\mathcal{M}$, observations, background covariance matrix, observation covariance matrices, window size $T$, total steps $L$

\State $t \gets 0$ \Comment{Initialize the time stamp}

\For{$step$ from $0$ to $L$} 
    \State $\mathbf{x}_t^a \gets \arg \min_{\mathbf{x}} J\left(\mathbf{x}\middle|\mathbf{x}_t^b, \{\mathbf{y}_\tau\}_{\tau=t}^{t+T-1}\right)$ \Comment{Solve 4DVar to obtain the analysis field.}
    \State $\mathbf{x}_{t+T}^b \gets \mathcal{M}_{t\to t+T} (\mathbf{x}_t^a)$ \Comment{Do Forecasting to obtain the background field at next time step.}
    \State $t \gets t + T$ 
\EndFor

\end{algorithmic}
\end{algorithm}  



\begin{figure}[htpb]
\centering
\includegraphics[width=1.0\linewidth]{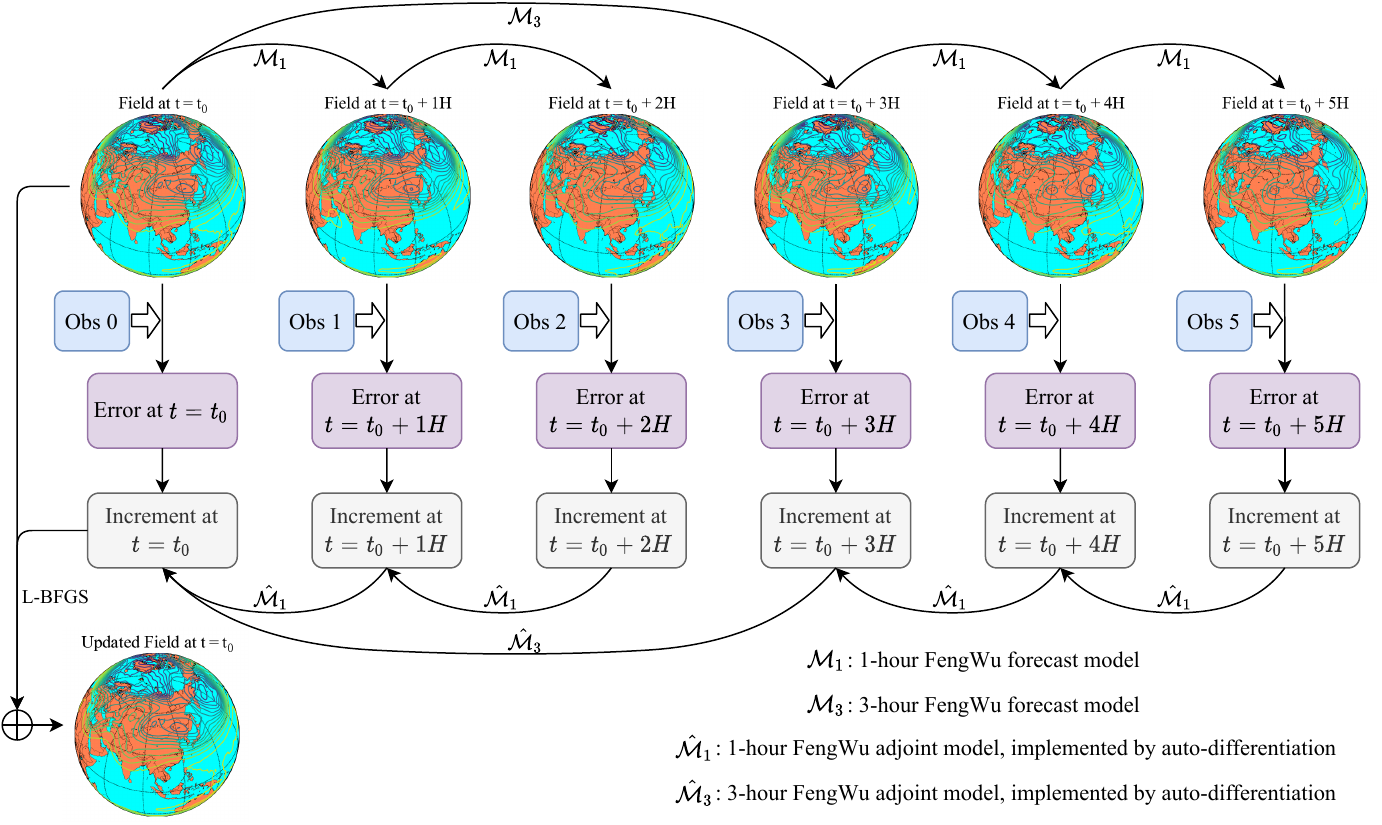}
\caption{\textbf{Schematic diagram of implementing 4DVar on the FengWu forecasting model (corresponding to one step optimization at time $t=t_0$).} The temporal aggregation strategy is represented by $\mathcal{M}_3$ and $\hat{\mathcal{M}}_3$ in the figure, enabling more direct and efficient propagation of gradients. The auto-differentiation strategy is represented by $\hat{\mathcal{M}}_1$ and $\hat{\mathcal{M}}_3$ in the figure and is automatically implemented by PyTorch, requiring no manual code writing.}
\label{fig:framework}
\end{figure}

\section{Integrating the AI Forecasting Model with 4DVar}

Assuming that we have an trained AI surrogate for the physical model, also denoted as $\mathcal{M}$ for simplicity, if we replace the physical model with its AI surrogate, the Bayesian formulation of the 4DVar objective function ($\ref{equ:4dvar}$) theoretically remains applicable. However, the subtle differences between the AI surrogate model and the physical model bring some difficulties in the construction and optimization of the objective function. This section focuses on discussing these difficulties and presenting the corresponding solutions. Our implementation structure is demonstrated in Figure~\ref{fig:framework}.

\subsection{Auto-Differentiation for Optimization}

For a general AI model $\mathcal{M}$, explicitly obtaining its adjoint is also not a straightforward task. Nevertheless, we can bypass this challenge and directly compute the gradient $\frac{\partial J}{\partial \mathbf{x}_0}$ by leveraging auto-differentiation~\citep{paszke2017automatic}.

Specifically, we can construct a new "neural network" with $\mathbf{x}_0$ (the analysis field to be optimized) as the input, and use this "neural network" to calculate the loss $J(\mathbf{x}_0)$. Noting that $\mathcal{M}$ is a differentiable AI model and other calculations like matrix multiplication and addition are also differentiable, if we fix the parameters of the AI model $\mathcal{M}$ and regard the input $\mathbf{x}_0$ as the "parameters" of our "neural network", the back-propagation algorithm can be done through auto-differentiation and thereafter the gradient $\frac{\partial J}{\partial \mathbf{x}_0}$ can be directly obtained.

Through this technique, we hand over the development of adjoint models, which requires a lot of engineering work, to the auto-differentiation function of the neural network. We add it here that although in our work no adjoint model is explicitly constructed when applying auto-differentiation to calculate $\frac{\partial J}{\partial \mathbf{x}_0}$, auto-differentiation has the capacity of explicitly constructing the adjoint model of an AI model, which is proved in Appendix~\ref{subsec:adjoint}.

To rigorously demonstrate that auto-differentiation can replace traditional adjoint-based gradient computation methods, we also need to establish the equivalence in the gradient computation processes of these two methods. This is crucial because numerical instability may lead to different results with different computation methods for highly nonlinear functions. We provide the proof of this equivalence in Appendix~\ref{subsec:equivalence}.

After computing the gradient, we utilize the L-BFGS optimizer from PyTorch~\citep{paszke2019pytorch} to optimize the objective function.

\subsection{Temporal Aggregation Strategy}

In the 4DVar objective function, $\left(\mathcal{H}\left(\mathbf{x}_\tau\right) - \mathbf{y}_\tau\right)^T \mathbf{R}_\tau^{-1} \left(\mathcal{H}\left(\mathbf{x}_\tau\right) - \mathbf{y}_\tau\right)$ measures how well the evolved future states align with the observations. To calculate $\mathbf{x}_\tau$, we need to integrate the physical state from time $0$ to time $\tau$. It requires that the physical model is relatively accurate; otherwise, additional error will be introduced and the observations further away from the initial time may not be well assimilated. For traditional numerical models, the implementation is straightforward, involving continuous integration of the model. Denote $\mathcal{M}_1$ the integration over one single time unit, then in traditional 4DVar, the objective function can be expressed as:
\begin{equation}
    \begin{aligned}
        J(\mathbf{x}_0) &= \frac12 \left(\mathbf{x}_0 - \mathbf{x}^b\right)^\text{T} \mathbf{B}^{-1} \left(\mathbf{x}_0 - \mathbf{x}^b\right) + \frac12 \sum_{\tau=0}^{T-1} \left(\mathcal{H}\left(\mathcal{M}_1^\tau(\mathbf{x}_0)\right) - \mathbf{y}_\tau\right)^\text{T} \mathbf{R}_\tau^{-1} \left(\mathcal{H}\left(\mathcal{M}_1^\tau(\mathbf{x}_0)\right) - \mathbf{y}_\tau\right),
    \end{aligned}
    \label{equ:tra-4dvar}
\end{equation}
where $\mathcal{M}_1^\tau = \mathcal{M}_1\circ\mathcal{M}_1\circ\cdots\circ\mathcal{M}_1$ (totally $\tau$ times).

In the case of an AI forecasting model, we can imitate the traditional method and train a neural network to integrate over a single time unit. Subsequently, the physical field can be obtained up to time $\tau$ by autonomously regressing the network for $\tau$ steps. This approach is theoretically correct, but assimilation accuracy may be hampered due to the characteristics of AI models. The issue lies in the current capabilities of AI forecasting models. When we train a neural network to integrate over a time unit (e.g., 1 hour) and achieve a level of accuracy comparable to traditional models, the subsequent autoregressive iteration of this network (e.g., 6 times) often results in errors that are higher than those of traditional models forecasting for the same duration (e.g., 6 hours). Moreover, this discrepancy becomes more pronounced with shorter time units and more iterations.

Here, we first explore the temporal aggregation strategy ~\citep{bi2023accurate} in the formulation of the 4DVar objective function to address this issue. Following the operational settings in ECMWF, we choose a 6-hour assimilation window and a 1-hour observation interval. In this setting, We train two AI surrogate models for the 4DVar algorithm: the first is used for 1-hour prediction, denoted by $\mathcal{M}_1$ and the second for 3-hour prediction, denoted by $\mathcal{M}_3$. $J(\mathbf{x}_0)$ can be formulated as
\begin{equation}
\begin{aligned}
J(\mathbf{x}_0) &= \frac12 \left(\mathbf{x}_0 - \mathbf{x}^b\right)^\text{T} \mathbf{B}^{-1} \left(\mathbf{x}_0 - \mathbf{x}^b\right) \\
&+ \frac12 \left(\mathcal{H}\left(\mathbf{x}_0\right) - \mathbf{y}_0\right)^\text{T} \mathbf{R}_0^{-1} \left(\mathcal{H}\left(\mathbf{x}_0\right) - \mathbf{y}_0\right) \\
&+ \frac12 \left(\mathcal{H}\left(\mathcal{M}_1(\mathbf{x}_0)\right) - \mathbf{y}_1\right)^\text{T} \mathbf{R}_1^{-1} \left(\mathcal{H}\left(\mathcal{M}_1(\mathbf{x}_0)\right) - \mathbf{y}_1\right) \\
&+ \frac12 \left(\mathcal{H}\left(\mathcal{M}_1\circ\mathcal{M}_1(\mathbf{x}_0)\right) - \mathbf{y}_2\right)^\text{T} \mathbf{R}_2^{-1} \left(\mathcal{H}\left(\mathcal{M}_1\circ\mathcal{M}_1(\mathbf{x}_0)\right) - \mathbf{y}_2\right) \\
&+ \frac12 \left(\mathcal{H}\left(\mathcal{M}_3(\mathbf{x}_0)\right) - \mathbf{y}_3\right)^\text{T} \mathbf{R}_3^{-1} \left(\mathcal{H}\left(\mathcal{M}_3(\mathbf{x}_0)\right) - \mathbf{y}_3\right) \\
&+ \frac12 \left(\mathcal{H}\left(\mathcal{M}_1\circ\mathcal{M}_3(\mathbf{x}_0)\right) - \mathbf{y}_4\right)^\text{T} \mathbf{R}_4^{-1} \left(\mathcal{H}\left(\mathcal{M}_1\circ\mathcal{M}_3(\mathbf{x}_0)\right) - \mathbf{y}_4\right) \\
&+ \frac12 \left(\mathcal{H}\left(\mathcal{M}_1\circ\mathcal{M}_1\circ\mathcal{M}_3(\mathbf{x}_0)\right) - \mathbf{y}_5\right)^\text{T} \mathbf{R}_5^{-1} \left(\mathcal{H}\left(\mathcal{M}_1\circ\mathcal{M}_1\circ\mathcal{M}_3(\mathbf{x}_0)\right) - \mathbf{y}_5\right) \\
\label{equ:4dvar-roll}
\end{aligned}
\end{equation}
By adopting this strategy, the accumulated errors during forward integration in assimilation can be reduced. Taking observations assimilated 5 hours away from the background field as an example, if we train a surrogate model for only 1 hour, our estimate of the background field at 5 hours is obtained by five consecutive 1-hour forward integrations from the initial state. On the other hand, if we employ the temporal aggregation strategy, this estimate is derived from two 1-hour integrations and one 3-hour integration. For an AI model, the errors from five 1-hour integrations are higher than those from two 1-hour integrations and one 3-hour integration. Therefore, the latter provides a more accurate estimate of the background field at 5 hours. In 4DVar algorithm, smaller model errors lead to higher assimilation accuracy. Hence, this strategy can theoretically enhance assimilation accuracy.

When engaged in cyclic forecasting, corresponding to $\mathbf{x}_{t+T}^b \gets \mathcal{M}_{t\to t+T} (\mathbf{x}_t^a)$ as outlined in Algorithm~\ref{alg:cyclic}, we introduce an additional 6-hour integration model, denoted as $\mathcal{M}_6$. The forecast process can be expressed as:
\begin{equation}
    \mathbf{x}_{t+6}^b = \mathcal{M}_6 (\mathbf{x}_t^a).
\label{equ:6hour-forecast}
\end{equation}
The purpose of training an additional 6-hour forecasting model is to reduce prediction errors. Without this dedicated 6-hour model, forecasting would require nesting the 3-hour and 1-hour forecasting models, leading to higher errors during prediction. We will further explore the replay buffer mechanism proposed in FengWu~\cite{chen2023fengwu} later, which can surrogate the physical model with only one neural network.

\subsection{Model Error Variance Term}

In order to further take model errors into account, we adjust the objective function by adding model error variance to it~\citep{howes2017accounting}. Assume that the error of the AI model (perhaps aggregated model) for integrating $i$ steps follows a Gaussian distribution $\mathcal{N}(\mathbf{0}, \mathbf{Q}_\tau)$. Then through Bayes' theorem, we can re-derive the objective function. We let readers refer to Appendix~\ref{subsec:model-error} for the detailed derivation. The form of the new objective function is almost identical to the original one, except that the model error covariance matrix needs to be added to the observation covariance matrix. That is,
\begin{equation}
    J(\mathbf{x}_0) = \frac12 \left(\mathbf{x}_0 - \mathbf{x}^b\right)^\text{T} \mathbf{B}^{-1} \left(\mathbf{x}_0 - \mathbf{x}^b\right) + \frac12 \sum_{\tau=0}^{T-1} \left(\mathcal{H}\left(\mathbf{x}_\tau\right) - \mathbf{y}_\tau\right)^\text{T} \left(\mathbf{R}_\tau + \mathbf{Q}_\tau \right)^{-1}  \left(\mathcal{H}\left(\mathbf{x}_\tau\right) - \mathbf{y}_\tau\right),
\end{equation}
where $\mathbf{x}_\tau$ is calculated with the temporal aggregation strategy.

\section{Results}
\label{sec:results}

\subsection{Experimental Setup}

\paragraph{FengWu forecasting model} In our experiments, the AI weather forecasting model we use is FengWu~\citep{chen2023fengwu}, a data-driven global medium-range weather forecasting model. We choose this model because it is a classic AI weather forecasting model known for its outstanding forecasting capabilities, extending beyond ten days. If our 4DVar implementation on this model proves effective, it implies the potential for generalization to other forecasting models with similar forecasting capabilities. 

We simulate five atmospheric variables (each with 13 pressure levels) and four surface variables, a total of 69 predictands. In this paper, the atmospheric variables are geopotential (z), relative humidity (r), zonal component of wind (u), meridional component of wind (v), and air temperature (t), whose 13 sub-variables at different vertical level are presented by abbreviating their short name and pressure levels (e.g., z500 denotes the geopotential height at a pressure level of 500 hPa). The four surface variables are 2-meter temperature (t2m), 10-meter u wind component (u10), 10-meter v wind component (v10), and mean sea level pressure (msl). The spatial resolution is $128\times 256$. In order to realize the temporal aggregation strategy, we have trained 3 models using ERA5 dataset of year 1979-2015, including $\mathcal{M}_1$ for 1-hour forecasting, $\mathcal{M}_3$ for 3-hour forecasting, and $\mathcal{M}_6$ for 6-hour forecasting. 

\paragraph{Observation Setup} In this study, all experiments conducted are simulation experiments by simulating observations from the ERA5 dataset. To make the experiments as close as possible to real-world scenarios, we made some modifications to the ERA5 reanalysis field to generate simulated observations. First, we introduce a random mask to the reanalysis field to simulate the sparse distribution of observation stations in real scenarios. Additionally, we add noise to the reanalysis field to simulate measurement errors at observation stations in a real environment. Unless stated otherwise, the mask proportion in our experiments is 15\%, indicating that only 15\% of the locations have observations. The standard deviation of observation noise is 0.001 times the standard deviation of the variable distribution.

\paragraph{Background Covariance Matrix} Background covariance matrix is crucial for variational assimilation~\citep{kalnay2003atmospheric, fisher2003background}. Due to the large dimensionality of the physical field, which approaches one million ($69\times 128\times 256$), explicitly writing out the $\mathbf{B}$ matrix would result in an impractical matrix with dimensions of one million by one million, making it nearly impossible to tackle with~\citep{fisher2003background}. In operational assimilation systems, it is common to decompose the $\mathbf{B}$ matrix into block-diagonal matrices and then store these non-zero elements. A commonly used decomposition method involves breaking $\mathbf{B}$ into three modules: $\mathbf{B}=\mathbf{U}\mathbf{U}^\text{T}, \mathbf{U}=\mathbf{U}_p\mathbf{U}_v\mathbf{U}_h\mathbf{S}$~\citep{bannister2008review, barker2004three}, where $\mathbf{U}_p$ expresses the correlation between different physical variables, $\mathbf{U}_v$ expresses the correlation between different vertical layers,  $\mathbf{U}_h$ expresses the horizontal correlation, $\mathbf{S}$ corresponds to the variance. Since our study does not focus on the construction of $\mathbf{B}$, we only consider $\mathbf{U}_h$ and $\mathbf{S}$ for building $\mathbf{B}$ in this work. Specifically, the lagged NMC method is used for estimating $\mathbf{U}_h$~\citep{bannister2008review} and a control variable transform is employed for facilitating optimization~\citep{bannister2008review2, barker2004three}. In our implementation, we drew inspiration from the background error generation code (GEN\_BE 2.0) developed by NCAR for the WRF model~\citep{descombes2015generalized}. We let readers refer to \citet{descombes2015generalized} for the detailed formulation of $\mathbf{B}$.
 
\paragraph{Cyclic Forecasting Setup} The initial state for starting our cyclic forecasting system is obtained from ERA5 dataset. Specifically, in our experiments, we aim to initiate our assimilation system from 00:00 on January 1, 2018. To obtain the background field for this time, we start from the ERA5 reanalysis field at 18:00 on December 31, 2017, integrate it using the 6-hour forecasting model for 1 step, and use the resulting field as the background field $\mathbf{x}_0^b$. We run the assimilation system for one year, concluding its operation at 23:00 on December 31, 2018.

\paragraph{Evaluation Metrics} To evaluate the performance of our cyclic forecasting system, we regard the ERA5 dataset as the ground truth and compare the analysis fields sequence $\{\mathbf{x}_{lT}^a\}_{l=0}^L$ and the background fields sequence $\{\mathbf{x}_{lT}^b\}_{l=0}^L$ with it. The metrics we use are RMSE and Bias~\citep{rasp2020weatherbench}. 

\textit{RMSE} corresponds to the latitude-Weighted Root Mean Square Error. It is a statistical metric widely used to assess the accuracy of a model’s predictions across different latitudes. Denote $\hat{x}_{l, c, w, h}$ the predicted value of the $l$-th sample at channel $c$ (It can either be the surface variable or the atmospheric variable at a certain pressure level.), and $w$ and $h$ represents the indices for each grid along the latitude and longitude indices. Denote $x_{l, c, w, h}$ the target value. Then the RMSE at channel $c$ is defined as 
\begin{equation}
    \text{RMSE}(c) = \frac{1}{L}\sum_{l=1}^L \sqrt{\frac{1}{W\cdot H}\sum_{w=1}^W\sum_{h=1}^H W\cdot\frac{\cos{(\alpha_{w,h})}}{\sum_{w^\prime=1}^W \cos{(\alpha_{w^\prime,h})}}\left(x_{l,c,w,h} - \hat{x}_{l,c,w,h}\right)^2},
\label{equ:rmse}
\end{equation}
where $\alpha_{w,h}$ is the latitude of point $(w, h)$.

\textit{Bias} corresponds to the latitude-Weighted Bias. It is widely used to assess the systematic bias of a model. Following the denotation above, the Bias at channel $c$ is defined as
\begin{equation}
    \text{Bias}(c) = \frac{1}{L}\sum_{l=1}^L \frac{1}{W\cdot H}\sum_{w=1}^W\sum_{h=1}^H W\cdot\frac{\cos{(\alpha_{w,h})}}{\sum_{w^\prime=1}^W \cos{(\alpha_{w^\prime,h})}}\left(\hat{x}_{l,c,w,h} - x_{l,c,w,h}\right).
\label{equ:bias}
\end{equation}

\subsection{Analysis Field Evaluation}

\begin{figure}[htpb]
\centering
\includegraphics[width=1.0\linewidth]{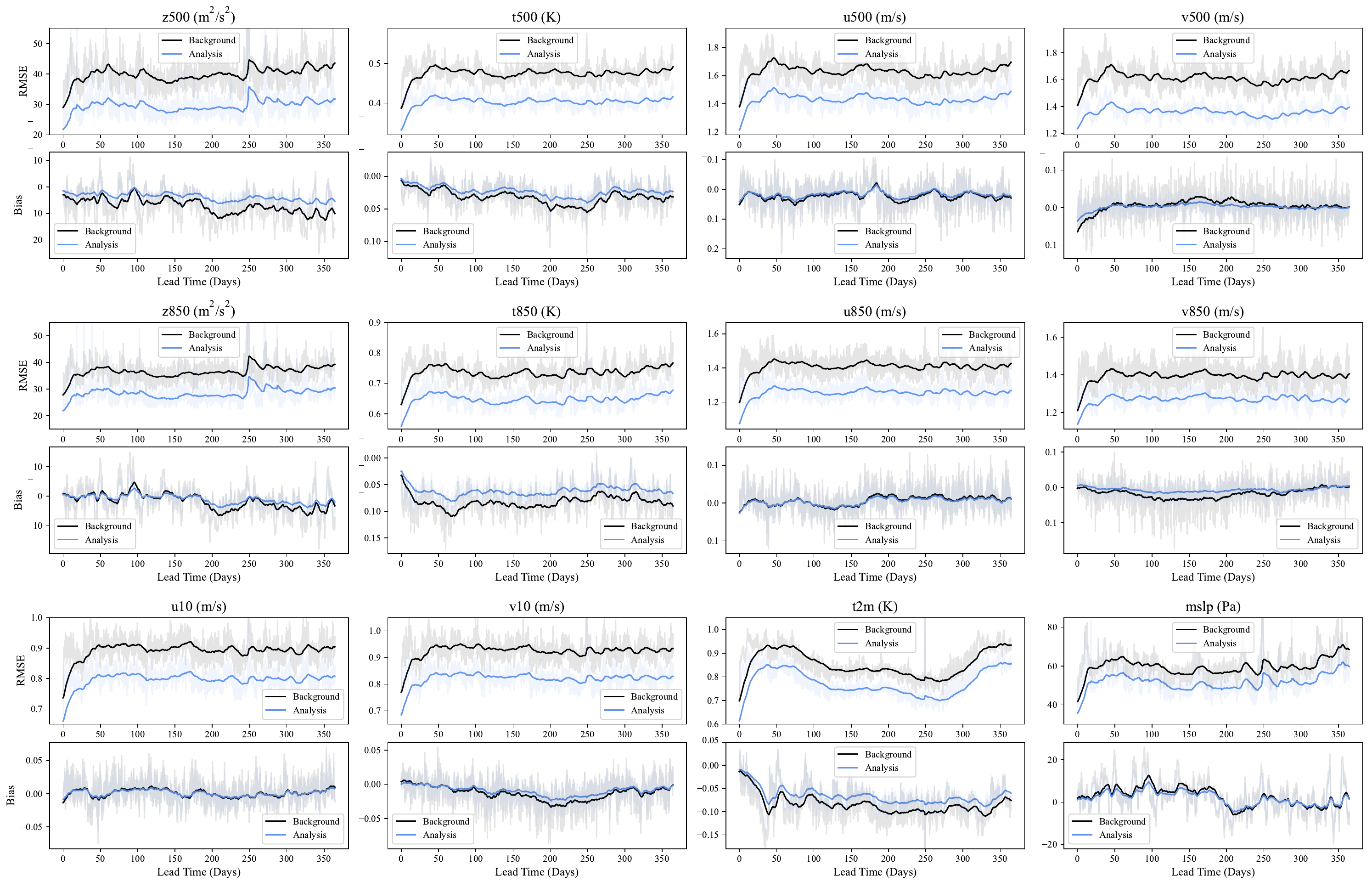}
\caption{\textbf{4DVar Assimilation Results on FengWu.} The x-axis in each sub-figure represents lead time, at a 6-hour interval over a one-year lead time. The y-axis represents the latitude-weighted RMSE and Bias defined in Equation~\ref{equ:rmse} and Equation~\ref{equ:bias}. The blue lines correspond to the analysis fields and the black lines correspond to the background fields.}
\label{fig:da-rmse-bias}
\end{figure}

Figure~\ref{fig:da-rmse-bias} demonstrates the cyclic forecasting results of 4DVar on FengWu. Four atmospheric variables (z, t, u, v) at two geopotential heights (500 hPa and 850 hPa) and four surface variables are reported. It can be found that on all reported variables, the RMSEs of the analysis fields are consistently smaller than those of the background fields. Take t500 as an example: the RMSE of the t500 channel of the background fields converges to around $0.48\,\mathrm{K}$, whereas that of the analysis fields converges to around $0.41\,\mathrm{K}$, indicating a decrease of $15\%$.
As for variables like temperature (t) and geopotential (z), the bias of the analysis fields is also reduced compared with the background fields. For other variables like wind speed (u, v) and mean sea level pressure (mslp), the Bias of the analysis fields does not show a significant improvement compared to the background field because, in the forecasts of these variables, the Fengwu model itself does not exhibit obvious systematic bias.

Additionally, for all reported variables, over the course of a year, the indicators of RMSE and Bias do not diverge with the progress of cyclic forecasting. Instead, they oscillate around a fixed value. This observation indicates that the 4DVar system based on the FengWu forecasting model, which we have constructed, can operate stably over the long term. After a long period of self-contained cyclic predictions, the RMSEs of the analysis fields of z500, t500, u500 and v500 are maintained around $30\,\mathrm{m^2/s^2}$, $0.4\,\mathrm{K}$, $1.41\,\mathrm{m/s}$, $1.39\,\mathrm{m/s}$, all of which indicate that our cyclic forecasting system has high predicting accuracy.



\subsection{Computational Cost}

Our data assimilation algorithm is implemented using auto-differentiation, and no additional neural network training is required once the forecasting model is trained. The primary computational expenses of the data assimilation algorithm arise from the calculations involving auto-differentiation and the updates to the analysis field using the PyTorch optimization algorithm. In our experiments, both auto-differentiation and gradient optimization updates are carried out on a single Nvidia A100 GPU, with an average runtime of 29.3 seconds for assimilating observations at 6-hour intervals.

\subsection{Ablation Study}

To comprehensively evaluate the assimilation system we have constructed, we carry out additional experiments to investigate the impact of the window size, temporal aggregation strategy, initial states, and $\mathbf{U}_h$ on assimilation performance. The outcomes of these experiments are detailed in this section.

\paragraph{Effect of Window Size}

We assess the impact of assimilating fewer observations by reducing the assimilation window size. Specifically, we shorten the assimilation window to 3 hours and 1 hour, respectively. It is important to note that the assimilation cycle remains at 6 hours, so the forecasting steps remain unchanged, with adjustments made to the 4DVar objective function as follows.
\begin{equation}
\begin{aligned}
J(\mathbf{x}_0) &= \frac12 \left(\mathbf{x}_0 - \mathbf{x}^b\right)^\text{T} \mathbf{B}^{-1} \left(\mathbf{x}_0 - \mathbf{x}^b\right) \\
&+ \frac12 \left(\mathcal{H}\left(\mathbf{x}_0\right) - \mathbf{y}_0\right)^\text{T} \mathbf{R}_0^{-1} \left(\mathcal{H}\left(\mathbf{x}_0\right) - \mathbf{y}_0\right) \\
&+ \frac12 \left(\mathcal{H}\left(\mathcal{M}_1(\mathbf{x}_0)\right) - \mathbf{y}_1\right)^\text{T} \mathbf{R}_1^{-1} \left(\mathcal{H}\left(\mathcal{M}_1(\mathbf{x}_0)\right) - \mathbf{y}_1\right) \\
&+ \frac12 \left(\mathcal{H}\left(\mathcal{M}_1\circ\mathcal{M}_1(\mathbf{x}_0)\right) - \mathbf{y}_2\right)^\text{T} \mathbf{R}_2^{-1} \left(\mathcal{H}\left(\mathcal{M}_1\circ\mathcal{M}_1(\mathbf{x}_0)\right) - \mathbf{y}_2\right) 
\label{equ:4dvar-win3}
\end{aligned}
\end{equation}
Equation~\ref{equ:4dvar-win3} is the objective functions of a 3-hour window 4DVar. 
\begin{equation}
J(\mathbf{x}_0) = \frac12 \left(\mathbf{x}_0 - \mathbf{x}^b\right)^\text{T} \mathbf{B}^{-1} \left(\mathbf{x}_0 - \mathbf{x}^b\right) + \frac12 \left(\mathcal{H}\left(\mathbf{x}_0\right) - \mathbf{y}_0\right)^\text{T} \mathbf{R}_0^{-1} \left(\mathcal{H}\left(\mathbf{x}_0\right) - \mathbf{y}_0\right) 
\label{equ:4dvar-win1}
\end{equation}
Equation~\ref{equ:4dvar-win1} is the objective functions of a 1-hour window 4DVar, which is reduced to the 3D-Var algorithm.

\begin{figure}[htpb]
\centering
\includegraphics[width=1.0\linewidth]{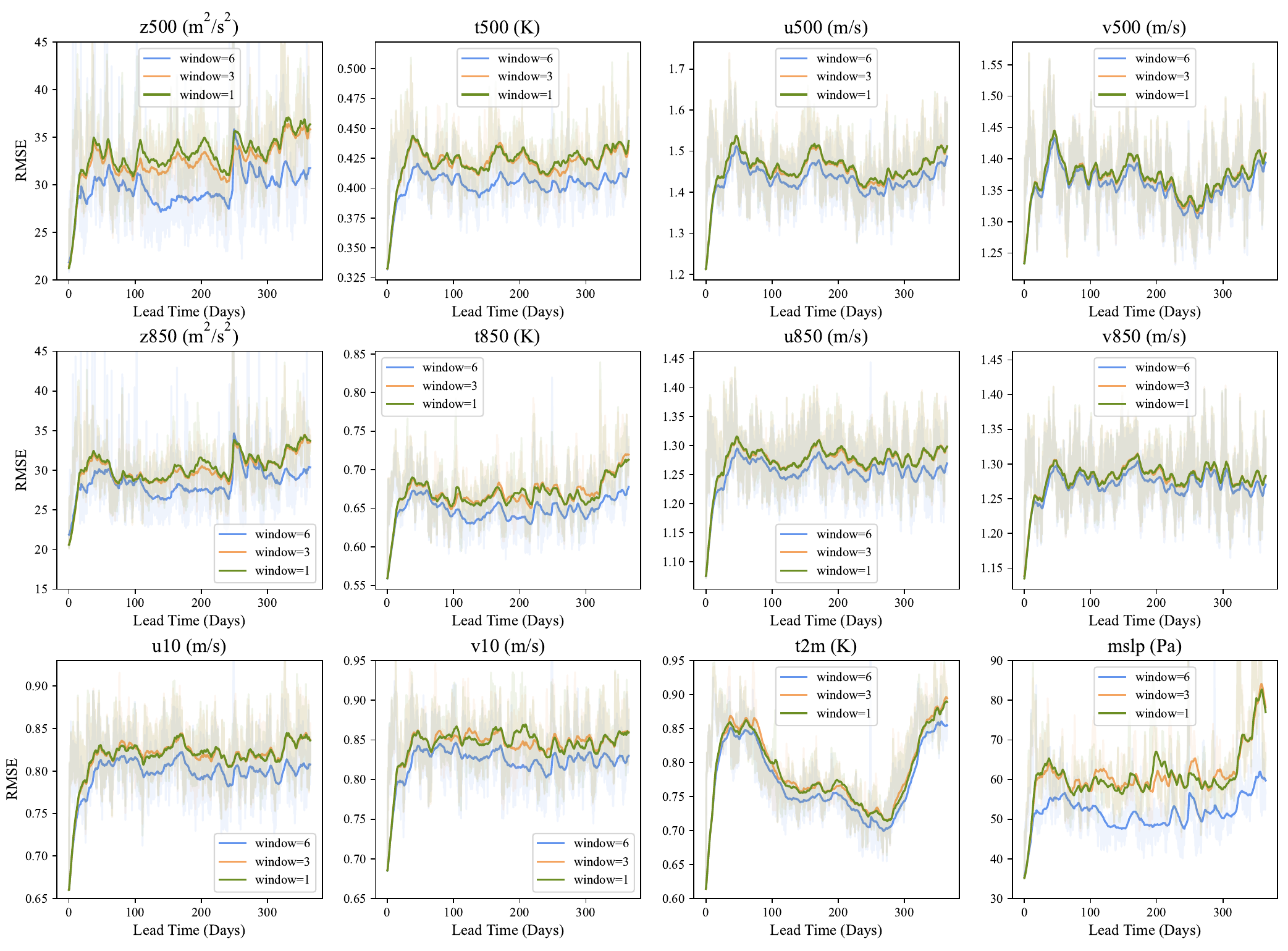}
\caption{\textbf{Effect of Window Size.} The x-axis in each sub-figure represents lead time, at a 6-hour interval over a one-year lead time. The y-axis represents the latitude-weighted RMSE defined in Equation~\ref{equ:rmse}. All the reults on reported on analysis fields. The blue, orange, green lines correspond to the 6-hour, 3-hour, and 1-hour windows, respectively.}
\label{fig:win_size}
\end{figure}

The RMSE of the analysis fields are reported in Figure~\ref{fig:win_size}. The results reveal that as the window size increases, the RMSE decreases, signifying an improved assimilation performance. This aligns with the intuitive understanding that a larger window size allows for more observations, enhancing the assimilation process and leading to better results.

\paragraph{Effect of Temporal Aggregation Strategy}

We conduct two controlled experiments to test the relevance of the temporal aggregation strategy. 

In the first experiment, we adopt the formulation in Equation~\ref{equ:tra-4dvar} and use the 1-hour model $\mathcal{M}_1$ only to calculate $\mathbf{x}_\tau$ in the 4DVar objective function. 

\begin{figure}[htpb]
\centering
\includegraphics[width=1.0\linewidth]{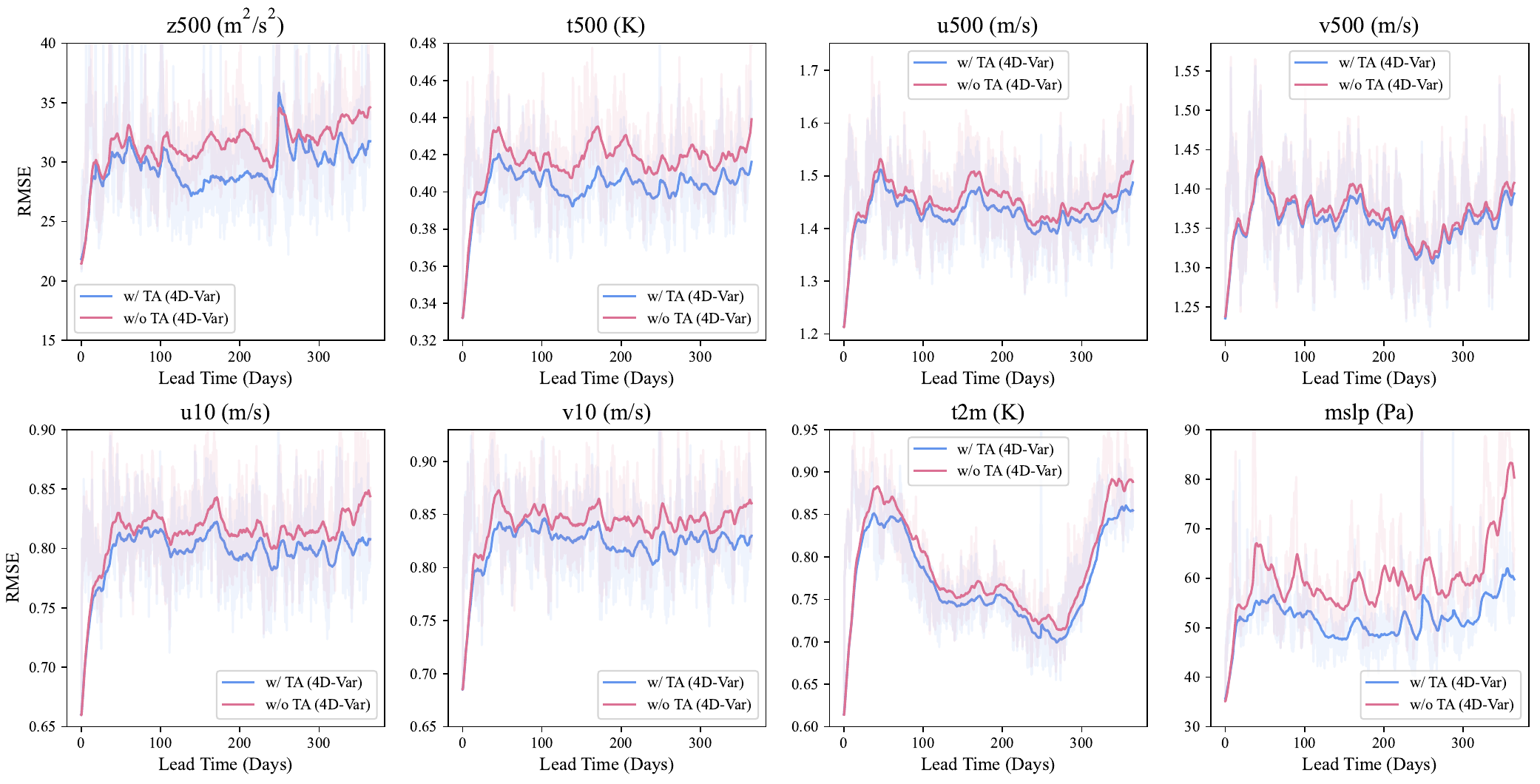}
\caption{\textbf{Effect of Temporal Aggregation Strategy (during 4DVar).} The label "w/ TA (4DVar)" corresponds to the experiment in which the temporal aggregation strategy is applied; the label "w/o TA (4DVar)" corresponds to that in which the temporal aggregation strategy is removed.}
\label{fig:tmp-agg}
\end{figure}

\begin{figure}[htpb]
\centering
\includegraphics[width=1.0\linewidth]{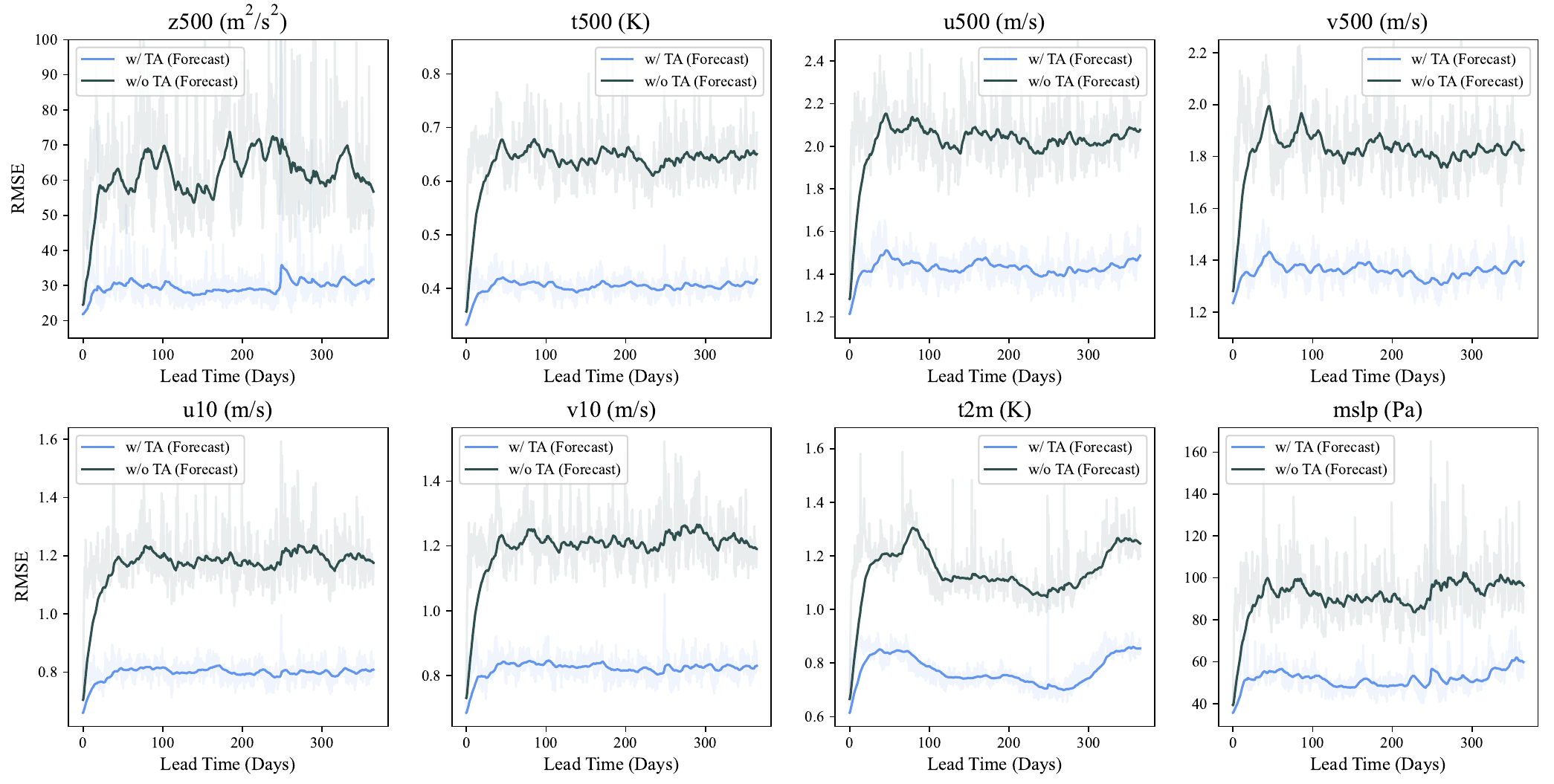}
\caption{\textbf{Effect of Temporal Aggregation Strategy (during Forecast).} The label "w/ TA (Forecast)" corresponds to the experiment in which the forecast is performed by a 6-hour forecasting model; the label "w/o TA (4DVar)" corresponds to that in which the forecast is performed by a combination of one 3-hour and three 1-hour forecasting models.}
\label{fig:forecast-strategy}
\end{figure}

As shown in Figure~\ref{fig:tmp-agg}, across all reported variables, the RMSE diminishes when the 3-hour model is utilized for computing the 4DVar objective function. This observation affirms that incorporating the temporal aggregation strategy is advantageous in reducing model error and improving the assimilation of observations.

In the second experiment, we change the forecasting setting. The original forecast process is performed with a 6-hour FengWu model $\mathcal{M}_6$, as expressed in Equation~\ref{equ:6hour-forecast}. In the new setting, the forecast process is performed with 1-hour and 3-hour FengWu models. That is,
\begin{equation}
    \mathbf{x}_{t+6}^b = \mathcal{M}_1 \circ \mathcal{M}_1 \circ \mathcal{M}_1 \circ \mathcal{M}_3 (\mathbf{x}_t^a).
\end{equation}
We conduct this experiment to prove that training a 6-hour FengWu model is indispensable for cyclic forecasting. 

It is shown in Figure~\ref{fig:forecast-strategy} that the assimilation results with 6-hour forecasting model is significantly better. As for some of the variables like z500, the assimilation system using only the 1-hour and 3-hour forecasting models exhibits RMSE values that are even 2-3 times higher compared to the system using the 6-hour forecasting model.

The two control experiments above confirm that aggregating three temporal integration models (1-hour, 3-hour, and 6-hour) leads to improved assimilation results.

\paragraph{Effect of Initial States} We choose different initial states, corresponding to $\mathbf{x}_0^b$ in Algorithm~\ref{alg:cyclic}, to test the robustness of our system. In the original experiment, the initial background field state was obtained by selecting the ERA5 reanalysis field from 6 hours prior to the initial time and then integrating it for 1 step using the 6-hour FengWu forecasting model.

Now, we select reanalysis fields from earlier time steps to introduce additional initial background field errors. Specifically, we integrate the reanalysis field from 2 days prior for 8 steps, from 4 days prior for 16 steps, and from 6 days prior for 24 steps, using these resulting states as initial conditions for assimilation experiments. These experiments are denoted as "init-2", "init-3", and "init-4" respectively. The original version of the experiment is labeled as "init-1".

\begin{figure}[htpb]
\centering
\includegraphics[width=1.0\linewidth]{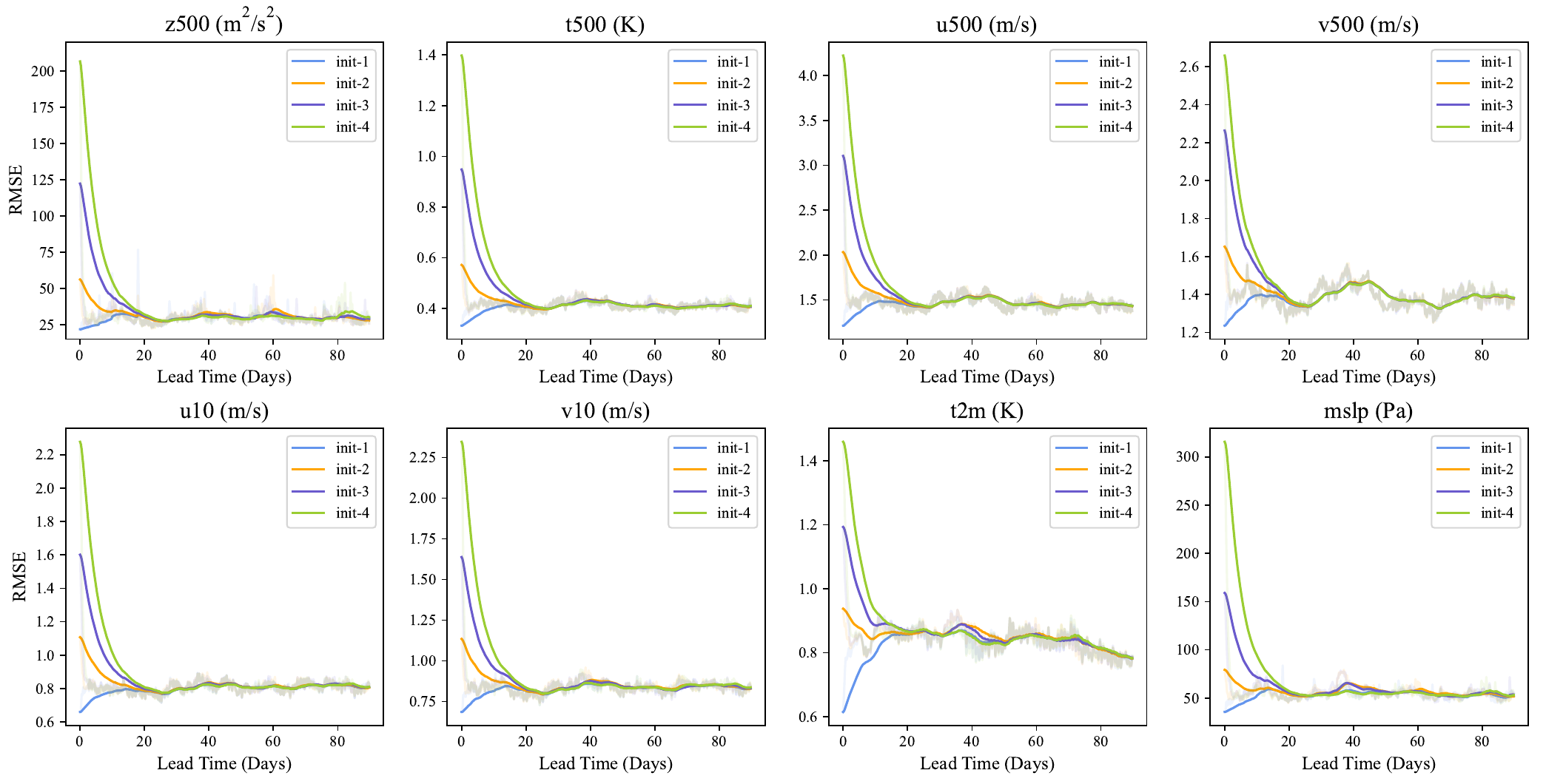}
\caption{\textbf{Effect of Initial States.} We let readers refer to the main text for the meanings of the labels.}
\label{fig:init-condition}
\end{figure}

The RMSE results on four atmospheric variables at geopotential height of 500hPa and four surface variables are reported in Figure~\ref{fig:init-condition}. It is shown that regardless of the magnitude of the initial state error, as we iterate the forecast for around 20 days, our data assimilation system consistently reduces the error to the same level as the original experiment setting. This indicates the successful correction of errors in the initial state through the assimilation of observations.

\begin{figure}[htpb]
\centering
\includegraphics[width=1.0\linewidth]{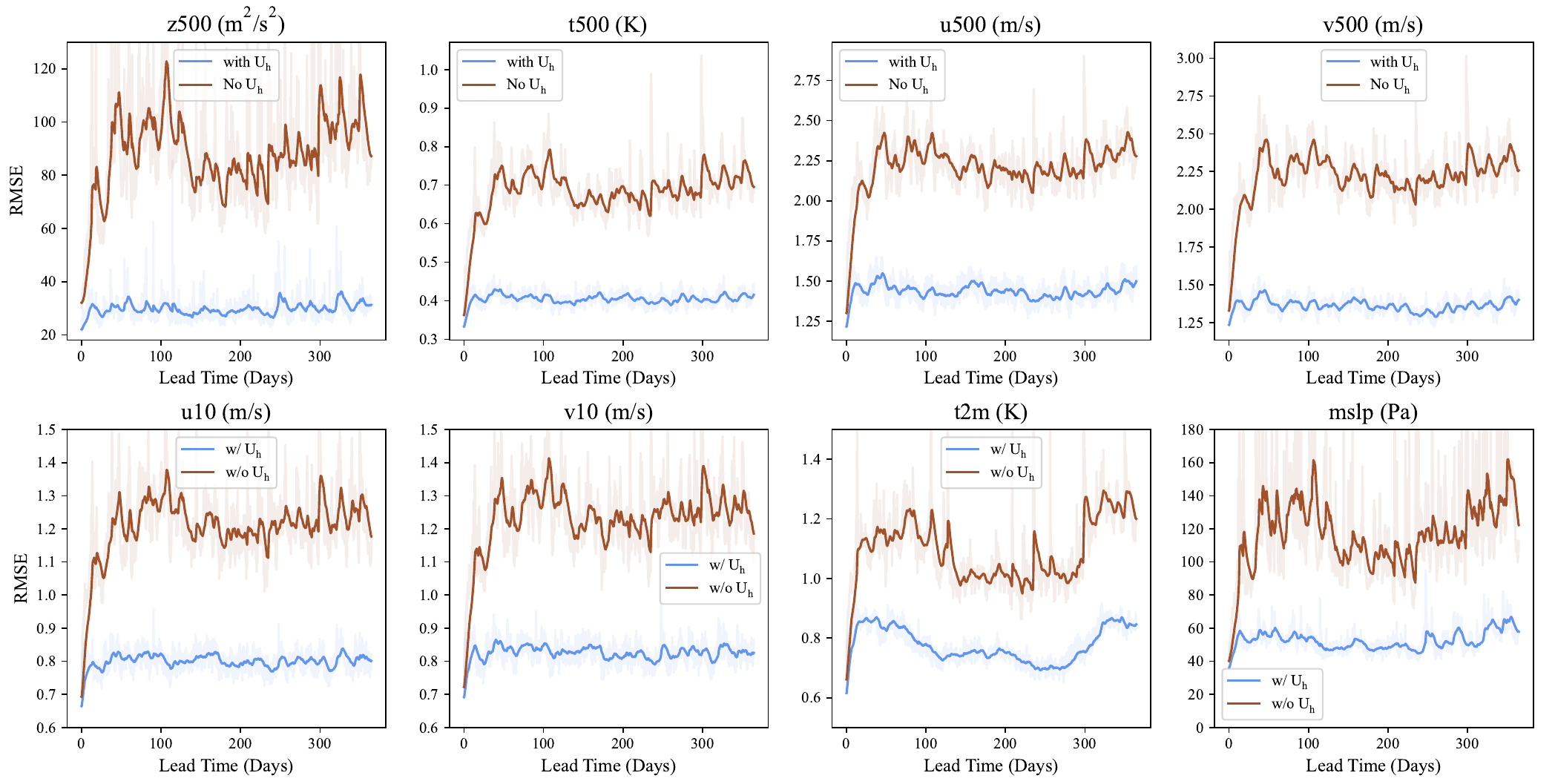}
\caption{\textbf{Effect of Horizontal Correlation in $\mathbf{B}$.} The label "w/ $\mathrm{U_h}$" corresponds to the experiment in which the horizontal correlation is included in the $\mathbf{B}$ matrix; the label "w/o $\mathrm{U_h}$" corresponds to that in which the horizontal correlation is excluded.}
\label{fig:uh-test}
\end{figure}

\paragraph{Effect of Horizontal Correlation in $\mathbf{B}$} In this experiment, we remove the horizontal correlation term $\mathbf{U}_h$ in the $\mathbf{B}$ matrix and test its cyclic forecasting skills. The results are demonstrated in Figure~\ref{fig:uh-test}, which indicates that eliminating the horizontal correlation also results in a noticeable increase in forecasting errors. For crucial variables such as z500, the RMSE experiences an approximately threefold rise. 


\section{Conclusions and Discussions}
\label{sec:conclusions}

In this paper, we propose an AI-based cyclic weather forecasting system named FengWu-4DVar, in which the advanced global AI weather forecasting model, FengWu, is coupled with the well-known assimilation algorithm, 4DVar, to realize self-contained iterative predictions. To address the complexity of constructing the adjoint model in 4DVar, we ingeniously utilize FengWu's auto-differentiation capabilities to solve the issue and avoid the development of the adjoint model. Additionally, we theoretically prove the equivalence of the auto-differentiation method and the adjoint-based method in the computation process of calculating gradients. To tackle the assimilation challenges of observations distant from the initial time in 4DVar, we employ the temporal aggregation strategy, which allows effective assimilation of observations even at distant time steps. 

We test this framework on simulated observations generated from the ERA5 dataset. Experimental results demonstrate that when coupled with 4DVar data assimilation, the FengWu weather forecasting model is capable of making stable cyclic predictions for at least one year. It is also found that FengWu-4DVar is efficient in cyclic forecasting. At a resolution of $128\times256$, the data assimilation system can assimilate observations within a 6-hour time window in less than half a minute on one Nvidia A100 GPU card. Furthermore, through sensitivity experiments, we investigate the impact of the assimilation window size, initial condition, and background field covariance matrix on assimilation accuracy. 

At the same time, we admit certain limitations in our current work. For instance, although FengWu-4DVar is able to make stable predictions during cycling forecasting, there is still room for improvement in the accuracy of the analysis field. In the current work, only horizontal correlations are considered for constructing the $\mathbf{B}$ matrix. In future work, we plan to incorporate vertical layer correlations and cross-variable correlations into the background error covariance matrix to further enhance assimilation accuracy. Additionally, our current work is conducted under simulated observational conditions, and the effectiveness in real observational scenarios is yet to be validated in future work.

\section*{Acknowledgements}

We express our gratitude for the utilization of the ERA5 dataset generously provided by the European Centre for Medium-Range Weather Forecasts (ECMWF). This study would not have been feasible without their commendable efforts in collecting, archiving, and disseminating the data.

We extend our appreciation to the Research Support, IT, and Infrastructure team based in the Shanghai AI Laboratory for providing us with computational resources and network support. 

This work is partially supported by the National Natural Science Foundation of China~(NO.U2242210) and the National Key R\&D Program of China~(NO.2022ZD0160101). 

We would also like to express our appreciation to Prof. Juanjuan Liu from the University of Chinese Academy of Science, Prof. Xiangjun Tian from the Institute of Tibetan Plateau Research, Chinese Academy of Sciences, Mr. Kun Chen and Mr. Hao Chen from Shanghai Artificial Intelligence Laboratory for their help and valuable discussions during the conduction of this research, which significantly contribute to improving the quality of this work. We express gratitude for their contributions and recognize their significant role in the development of this work.

\bibliographystyle{unsrtnat}
\bibliography{references}  

\newpage

\appendix

\section{Mathematical Details}

\subsection{Constructing the Adjoint Model through auto-differentiation}
\label{subsec:adjoint}

Assuming that $\mathcal{M}: \mathbb{R}^d\to\mathbb{R}^d$ is a forecasting model (the physical field flattened into a one-dimensional vector), the adjoint model of $\mathcal{M}$ at state $\mathbf{x}_0$ is defined as $\left(\frac{\partial\mathcal{M}(\mathbf{x})}{\partial \mathbf{x}}\middle|_{\mathbf{x}=\mathbf{x}_0}\right)^\text{T}$. It is a $d\times d$ matrix. 

\begin{figure}[htpb]
\centering
\includegraphics[width=0.6\linewidth]{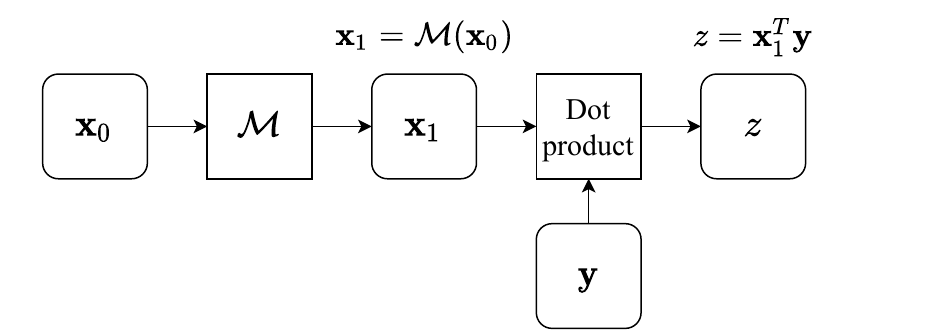}
\caption{\textbf{"Neural Network" for calculating the adjoint model.}}
\label{fig:CG-ad}
\end{figure}

Suppose $\mathbf{y}\in \mathbb{R}^{d\times d}$ is another arbitrary $d$-dimensional vector. Then we can build the following "neural network", as shown in Figure~\ref{fig:CG-ad}. It only consists of two steps. In the first step, $\mathbf{x}_0$ is fed into the forecasting model and the predicted state $\mathbf{x}_1$ is obtained. In the second step, we do a dot product between $\mathbf{x}_1$ and $\mathbf{y}$ and produce a scalar $z$ as the output.

Through auto-differentiation, we can obtain the gradient at node $\mathbf{x}_0$, that is $\frac{\partial z}{\partial \mathbf{x}_0}$. On the other hand, we can do the computational graph manually, and find out what the gradient stands for:
\begin{equation}
\begin{aligned}
    \frac{\partial z}{\partial \mathbf{x}_1} &= \mathbf{y} \\
    \frac{\partial z}{\partial \mathbf{x}_0} &= \left(\frac{\partial \mathbf{x}_1}{\partial \mathbf{x}_0}\right)^T \frac{\partial z}{\partial \mathbf{x}_1} = \left(\left.\frac{\partial \mathcal{M}(\mathbf{x})}{\partial \mathbf{x}} \right|_{\mathbf{x}=\mathbf{x}_0}\right)^T \mathbf{y}
\end{aligned}
\label{equ:CG-ad}
\end{equation}
According to Equation~\ref{equ:CG-ad}, it can be found that the gradient at node $\mathbf{x}_0$ precisely represents the result of the adjoint model (defined at $\mathbf{x}_0$) acting on $\mathbf{y}$. Since both $\mathbf{x}_0$ and $\mathbf{y}$ are arbitrary, through this approach, we can calculate the results of the adjoint model, defined at any point, acting on any vectors. This concludes the proof that the adjoint of any differentiable forecasting model can be constructed through auto-differentiation.

\subsection{Equivalence Between Two Optimization Methods}
\label{subsec:equivalence}

\begin{figure}[htpb]
\centering
\includegraphics[width=1.0\linewidth]{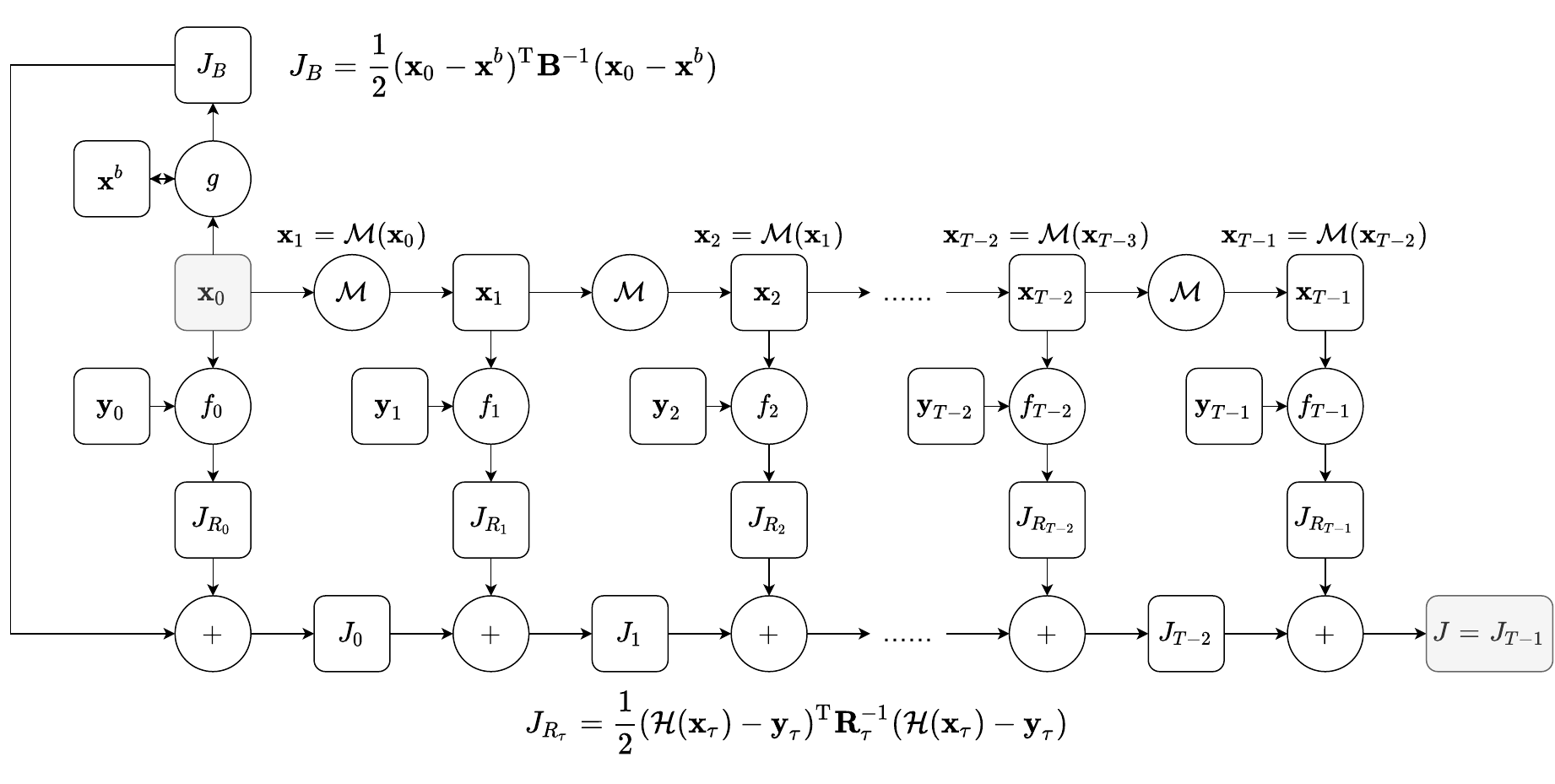}
\caption{\textbf{"Neural Network" for calculating the 4DVar objective function.}}
\label{fig:CG-4dvar}
\end{figure}

Define $g(\mathbf{x}_0, \mathbf{x}^b) = \frac{1}{2}(\mathbf{x}_0-\mathbf{x}^b)^\text{T}\mathbf{B}^{-1}(\mathbf{x}_0-\mathbf{x}^b)$, $f_i(\mathbf{x}, \mathbf{y})=\frac{1}{2}(\mathcal{H}(\mathbf{x})-\mathbf{y})^\text{T}\mathbf{R}_i^{-1}(\mathcal{H}(\mathbf{x})-\mathbf{y})$. Let $\mathcal{M}_{0\to 1}=\mathcal{M}_{1\to 2}=\cdots=\mathcal{M}_{T-2\to T-1}=\mathcal{M}$. Then the computational graph of calculating the 4DVar objective function can be constructed, as shown in Figure~\ref{fig:CG-4dvar}.
For simplicity, Denote $\mathbf{H}_\tau = \left.\frac{\partial \mathcal{H}(\mathbf{x})}{\partial \mathbf{x}} \right|_{\mathbf{x}=\mathbf{x}_\tau}$ and $\mathbf{M}_\tau = \left.\frac{\partial \mathcal{M}(\mathbf{x})}{\partial \mathbf{x}} \right|_{\mathbf{x}=\mathbf{x}_\tau} = \frac{\partial \mathbf{x}_{\tau+1}}{\partial \mathbf{x}_\tau}$. We can simulate the computational graph back-propagation process and calculate the gradient sequentially:
\begin{equation}
\begin{aligned}
    \frac{\partial J}{\partial \mathbf{x}_{T-1}} &= \frac{\partial J_{R_{T-1}}}{\partial \mathbf{x}_{T-1}} \frac{\partial J}{\partial J_{R_{T-1}}} = \frac{\partial J_{R_{T-1}}}{\partial \mathbf{x}_{T-1}} = \mathbf{H}^\text{T}_{T-1} \mathbf{R}_{T-1}^{-1} (\mathcal{H}(\mathbf{x}_{T-1}) - \mathbf{y}_{T-1}), \\
    \frac{\partial J}{\partial \mathbf{x}_{T-2}} &= \frac{\partial J_{R_{T-2}}}{\partial \mathbf{x}_{T-2}} \frac{\partial J}{\partial J_{R_{T-2}}} + \left(\frac{\partial \mathbf{x}_{T-1}}{\partial \mathbf{x}_{T-2}}\right)^\text{T} \frac{\partial J}{\partial \mathbf{x}_{T-1}}  = \frac{\partial J_{R_{T-2}}}{\partial \mathbf{x}_{T-2}} + \left(\frac{\partial \mathbf{x}_{T-1}}{\partial \mathbf{x}_{T-2}}\right)^\text{T} \frac{\partial J}{\partial \mathbf{x}_{T-1}} \\
    &= \mathbf{H}^T_{T-2} \mathbf{R}_{T-2}^{-1} (\mathcal{H}(\mathbf{x}_{T-2}) - \mathbf{y}_{T-2}) + \mathbf{M}_{T-2}^\text{T} \frac{\partial J}{\partial \mathbf{x}_{T-1}},
\end{aligned}
\end{equation}
Continuing in this manner, we can derive a general formula:
\begin{equation}
\begin{aligned}
    \frac{\partial J}{\partial \mathbf{x}_{\tau}} &= \frac{\partial J_{R_{\tau}}}{\partial \mathbf{x}_{\tau}} \frac{\partial L}{\partial J_{R_{\tau}}} + \left(\frac{\partial \mathbf{\tau}_{k+1}}{\partial \mathbf{x}_{\tau}}\right)^\text{T} \frac{\partial J}{\partial \mathbf{x}_{\tau+1}} = \frac{\partial J_{R_{\tau}}}{\partial \mathbf{x}_{\tau}} + \left(\frac{\partial \mathbf{x}_{\tau+1}}{\partial \mathbf{x}_{\tau}}\right)^\text{T} \frac{\partial J}{\partial \mathbf{x}_{\tau+1}} \\
    &= \mathbf{H}^\text{T}_{\tau} \mathbf{R}_{\tau}^{-1} (\mathcal{H}(\mathbf{x}_{\tau}) - \mathbf{y}_{\tau}) + \mathbf{M}_{\tau}^\text{T} \frac{\partial J}{\partial \mathbf{x}_{\tau+1}}.
\end{aligned}
\label{equ:4dvar-intermediate}
\end{equation}
This corresponds to the gradient stored at the intermediate node $\mathbf{x}_{\tau}$, which is calculated through auto-differentiation. 

Equation~\ref{equ:4dvar-intermediate} holds for $1 \leq \tau \leq T-2$; when $\tau=0$, an additional background term should be included:
\begin{equation}
    \frac{\partial J}{\partial \mathbf{x}_{0}} = \frac{\partial J_{R_{B}}}{\partial \mathbf{x}_{0}} \frac{\partial J}{\partial J_{B}} + \frac{\partial J_{R_{0}}}{\partial \mathbf{x}_{0}} \frac{\partial J}{\partial J_{R_{0}}} + \left(\frac{\partial \mathbf{x}_{1}}{\partial \mathbf{x}_{0}}\right)^\text{T} \frac{\partial J}{\partial \mathbf{x}_{1}} = \mathbf{B}^{-1}(\mathbf{x}_{0} - \mathbf{x}_{b}) + \mathbf{H}^\text{T}_{0} \mathbf{R}_{0}^{-1} (\mathcal{H}(\mathbf{x}_{0}) - \mathbf{y}_{0}) + \mathbf{M}_{0}^\text{T} \frac{\partial L}{\partial \mathbf{x}_{1}}.
\end{equation}
Up to this point, we have elucidated the mechanism through which auto-differentiation computes the gradient. By comparing this process with the methodology employed by traditional adjoint model-based methods for gradient calculation~\citep{rabier2003variational}, it becomes evident that these two processes are entirely identical.

\subsection{Derivation of the Model Error Term}
\label{subsec:model-error}

First, we assume that the model's errors follow a Gaussian distribution. It's worth noting that this assumption holds true only for linear models. For nonlinear models, the preservation of Gaussian distribution errors cannot be guaranteed during model integration. However, this does not prevent us from making such an approximation to make the problem manageable. Denote $\mathbf{Q}_\tau$ the variance of the error $\tau$-step integration model, $\mathcal{M}_{0\to\tau}$ or $\mathcal{M}_\tau$, then $\mathbf{x}_{\tau}|\mathbf{x}_0\sim \mathcal{N}(\mathcal{M}_\tau(\mathbf{x}_0), \mathbf{Q}_\tau)$. Since the observations at time $\tau$ also follows the Gaussian distribution $\mathbf{y}_\tau|\mathbf{x}_{\tau}\sim \mathcal{N}(\mathbf{x}_{\tau}, \mathbf{R}_\tau)$, through the compound rule of Gaussian distributions, $\mathbf{y}_{\tau}|\mathbf{x}_0$ also follows the Gaussian distribution:
\begin{equation}
    \mathbf{y}_{\tau}|\mathbf{x}_0 \sim \mathcal{N}(\mathcal{M}_\tau(\mathbf{x}_0), \mathbf{R}_\tau + \mathbf{Q}_\tau)
\end{equation}
According to Bayesian Theorem, we have
\begin{equation}
    \begin{aligned}
        \arg \max_{\mathbf{x}_0} p(\mathbf{x}_0|\mathbf{y}_0,\cdots,\mathbf{y}_{T-1}) = \arg \max_{\mathbf{x}_0} \frac{p(\mathbf{y}_0,\cdots,\mathbf{y}_{T-1}|\mathbf{x}_0)p(\mathbf{x}_0)}{p(\mathbf{y}_0,\cdots,\mathbf{y}_{T-1})} = \arg \max_{\mathbf{x}_0} p(\mathbf{y}_0,\cdots,\mathbf{y}_{T-1}|\mathbf{x}_0)p(\mathbf{x}_0).
    \end{aligned}
\end{equation}
Since observations at different time steps are independent, the above formula can be simplified as follows:
\begin{equation}
    \begin{aligned}
        &p(\mathbf{y}_0,\cdots,\mathbf{y}_{T-1}|\mathbf{x}_0)p(\mathbf{x}_0)
        = p(\mathbf{x}_0)\prod_{\tau=0}^{T-1}p(\mathbf{y}_\tau|\mathbf{x}_0)\\
        = &C \exp\left(-\frac12 \left(\mathbf{x}_0 - \mathbf{x}^b\right)^\text{T} \mathbf{B}^{-1} \left(\mathbf{x}_0 - \mathbf{x}^b\right)\right) \prod_{\tau=0}^{T-1} \exp\left(-\frac12 \left(\mathcal{M}_\tau(\mathbf{x}_0) - \mathbf{y}_\tau\right)^\text{T} (\mathbf{R}_\tau + \mathbf{Q}_\tau)^{-1} \left(\mathcal{M}_\tau(\mathbf{x}_0) - \mathbf{y}_\tau\right)\right),
    \end{aligned}
\end{equation}
where $C$ is a constant. Taking negative logarithm of this likelihood yields the objective function,
\begin{equation}
    J(\mathbf{x}_0) = \frac12 \left(\mathbf{x}_0 - \mathbf{x}^b\right)^\text{T} \mathbf{B}^{-1} \left(\mathbf{x}_0 - \mathbf{x}^b\right) + \frac12 \sum_{\tau=0}^{T-1} \left(\mathcal{M}_\tau(\mathbf{x}_0) - \mathbf{y}_\tau\right)^\text{T} (\mathbf{R}_\tau + \mathbf{Q}_\tau)^{-1} \left(\mathcal{M}_\tau(\mathbf{x}_0) - \mathbf{y}_\tau\right),
\end{equation}
which concludes the proof.


\end{document}